\documentclass[11pt,a4paper]{article}
\begin{document}
\def\rf#1{(\ref{eq:#1})}
\def\lab#1{\label{eq:#1}}
\def\nonu{\nonumber}
\def\br{\begin{eqnarray}}
\def\er{\end{eqnarray}}
\def\be{\begin{equation}}
\def\ee{\end{equation}}
\def\lb{\lbrack}
\def\rb{\rbrack}
\def\Blb{\Bigl\lbrack}
\def\Brb{\Bigr\rbrack}
\def\lcurl{\left\{}
\def\rcurl{\right\}}
\def\>{\rangle}              
\def\<{\langle}              
\def\({\left(}
\def\){\right)}
\def\[{\left[}
\def\]{\right]}
\def\v{\vert}                     
\def\bv{\bigm\vert}               
\def\tr{\mathop{\rm tr}}                  
\def\Tr{\mathop{\rm Tr}}                  
\def\pr{\prime}
\def\ra{\rightarrow}
\def\lra{\longrightarrow}
\def\f{\frac}
\def\grad{\nabla}
\def\ti{\tilde}
\def\wti{\widetilde}
\def\eq{\!\!\!\! &=& \!\!\!\! }
\def\jp{J_{+}}
\def\jm{J_{-}}
\def\j3{J_3}
\def\Ad{{\cal A}^{\dagger}}
\def\A{{\cal A}}

\def\half{\frac{1}{2}}
\def\id{i\partial_\phi}
\def\hm{h_{min}}

\def\a{\alpha}
\def\b{\beta}
\def\c{\chi}
\def\t{\theta}
\def\T{\Theta}

\newcommand{\vp}{\varphi}
\newcommand{\sech} {{\rm sech}}
\newcommand{\cosech} {{\rm cosech}}
\newcommand{\psib} {\bar{\psi}}
\newcommand{\cosec} {{\rm cosec}}

\title{investigation of the recurrence relations for the spheroidal
wave functions \footnote{ E-mail: tgh-2000@263.net,
tgh20080827@gmail.com, shuqzhong@gmail.com}}
\author{Guihua Tian$^{1,2}$,
\ Shuquan Zhong$^{1}$ \\
1.School of Science, Beijing University of Posts And
Telecommunications.
\\ Beijing 100876 China.\\
2.Department of Physics, University of Maryland, College Park, \\
Maryland 20742-4111 U.S.A}
\date{May 28, 2009}
\maketitle
\begin{abstract}
The perturbation method in supersymmetric quantum mechanics
(SUSYQM) is used to study the spheroidal wave functions'
recurrence relations, which are revealed by the shape-invariance
property of the super-potential. The super-potential is expanded
by the parameter $\a$ and could be gotten by approximation method.
Up to the first order, it has the shape-invariance property and
the excited spheroidal wave functions are gotten. Also, all the
first term eigenfunctions obtained are in closed form. They are
advantageous to investigating for involved physical problems
 of spheroidal wave function. \\
\textbf{PACs:11.30Pb; 04.25Nx; 04.70-s}
\end{abstract}

\section{Introduction of the spheroidal functions}

The spheroidal wave equations are extension of the ordinary
spherical wave equations. There are many fields where spheroidal
functions play important roles just as the spherical functions do.
So far, in comparison to simpler spherical special functions (the
associated Lengdre's functions ) their properties still are
difficult for study than their
counterpart\cite{flammer}-\cite{li}.

Their differential equations are
\begin{equation} \left[\frac{d}{dx}\left[(1-x^2) \frac{d}{d
x}\right]+E+ \a x^2
 -\frac{m^{2}}{1-x^2} \right]\Theta=0 x\in (-1,+1).\label{3}
\end{equation}
They only have one more term $\a X^2$ than the spherical ones (the
associated Lendgre's equations).

 This is a kind
of the Sturm-Liouville eigenvalue problem with the natural
conditions that $\Theta $ is finite at the boundaries $x=\pm 1$.
The parameter $E $ can only takes the values $E_0,\ E_1,\dots,
E_n,\dots$, which are called the eigenvalues of the problem, and
the corresponding solutions (the eigenfunctions) $\T_0,\T_1,\dots,
\T_n,\dots$   are called the spheroidal wave functions
\cite{flammer}-\cite{li}.

Under the condition $\a=0$, they reduce to the Spherical equation
 and  the solutions to the Sturm-Liouville eigenvalue problem are
the associated Legendre-functions $P_n^m(x)$(the spherical
functions) with the eigenvalues $E_n=n(n+1),\ n=m+1,\ m+2,\dots$.
Though the spheroidal wave equations are extension of the ordinary
spherical wave functions equations, the difference between this
two kinds of wave funtions are far greater than their
similarity\cite{flammer}.

\begin{enumerate}

\item{The spherical
 wave functions $P_n^m(x)$ have many good property such as
\begin{enumerate}
  \item{the Legendre-functions $P_n(x)=P_n^0(x)$ are polynomials;}
  \item{Back to the variable $\t$ with $x=\cos\t$, the recursion relation among the Legendre-functions could be written as:
  \br P_{n}(\cos\t)\propto \[-n\cos\t-\sin\t\f{d}{d\t}\]P_{n-1}(\cos\t),\
  n=1,2,3,\dots,\er
  so, all $P_n(\cos\t)$ could be deduced from the first or ground
function $P_0$ from the recursion relation }
  \item{all the associated
Legendre-functions $P_n^m(\cos\t)$ could be derived from $P_m^m$
by \br
P_n^m(\cos\t)=\[-n\cos\t-\sin\t\f{d}{d\t}\]P_{n-1}^m(\cos\t),\
  n=m+1,m+2,m+3,\dots,\label{p-l}.\er}
\end{enumerate}}
\item{On the contrary, very little are known about the properties of the spheroidal wave functions:
 \begin{enumerate} \item{ All spheroidal wave functions can
not be polynomials at all.}
\item{Whether or not there are the similar recurrence relations between them
  is not clear, at least they  have not been found upon to
  now.}\end{enumerate}}
\end{enumerate}

The recursion relations of the spherical wave functions, the
associated Legendre functions reflect the the factorization
properties of the Legendre equations\cite{Infeld}. This shows that
the Legendre equations are solvable.  The Super-symmetry quantum
mechanics tells that the roots of the solvable properties of
differential equations are the super-invariance of their
super-potential. In this paper, we testily investigate whether the
spheroidal differential equations have the similar
shape-invariance property. On accounting of the fact that their
differential equations are difficult to treat, we only can rely on
the approximate method of small parameter $\a$. The results may
make one happy: they have shape-invariance property upon to the
first term approximation.

\section{Brief review of the theory of SUSY in solving eigenvalue-problem }
   In recent years, supersymmetric quantum mechanics have attracted
tremendous attention for solvable potential problems. They not
only  provide clear insight into the factorization method of
Infeld and Hull \cite{Infeld}, but also make great improvement in
solving the differential equations. See reference \cite{Cooper}
for review on its development.

In supersymmetric quantum mechanics \cite{Cooper}-\cite{dutt}, the
Schr\"{o}dinger equation is \be H^-\psi^-=- {d^2 \psi^-\over dx^2
} + V^-(x) \psi^-(x)  = E_-\psi^- \ee with $\hbar=2m=1$. Here one
considers the case of unbroken supersymmetry which demands that
the ground state is nodeless with zero energy, and supposes that
the superpotential $W(x)$ is continuous and differentiable. The
super-potential $W(x)$ satisfies the equation
\begin{equation} \label{vpm}
V^{-}(x)=W^2(x) - W'(x).
\end{equation}
With the introducing the two operators \be \A = {d \over {dx}}
+W(x)~,~ \Ad=-{d \over {dx}} +W(x), \ee the corresponding
Hamiltonian $H^{-}$ have a factorized form \be \label{hm} H^-=\Ad
\A~. \ee The partner potential $V^{+}(x)$ of $V^-$ is determined
by the superpotential $W(x)$ as \be V^{+}(x)=W^2(x) + W'(x), \ee
and the corresponding Hamiltonian $H^+$ \be H^+\psi^+=- {d^2
\psi^+ \over dx^2 } + V^-(x) \psi^+(x)  = E_+ \psi^+ \ee  has a
factorized form \be H^+=\A \Ad~ .\ee

The Hamiltonians $H^+$ and $H^-$ have exactly the same eigenvalues
(or the same spectrum) except that $H^-$ has an additional zero
energy and the related eigenstate, that is,  \be \label{susy}
E^{(-)}_0 = 0~,~~~E^{(+)}_{n-1}\;=\;E^{(-)}_{n},~~~
\psi^{(+)}_{n-1} \propto \A \,\psi^{(-)}_{n}~~,~~~\Ad
\,\psi^{(+)}_{n} \propto \psi^{(-)}_{n+1}  ~~, \quad   n=1,2,
\ldots. \ee The pair of SUSY partner potentials $V^{\pm}(x)$ are
called shape invariant if they are similar in shape and differ
only in the parameters, that is
\begin{equation} \label{shape invariant} V^+(x;a_1) = V^-(x;a_2) +
R(a_1),
\end{equation}
where $a_1$ is a set of parameters, $a_2$ is a function of $a_1$
(say $a_2=f(a_1)$) and the remainder $R(a_1)$ is independent of
$x$. One can use the property of shape invariance to obtain the
analytic determination of energy eigenvalues and eigenfunctions
\cite{Cooper}-\cite{dutt}. Thus for an unbroken supersymmetry, the
eigenstates of the potential $V^-(x)$ are: \br && E_0^-=0,\ \
E_n^-=\sum_{k=1}^{n}R(a_k)\\
&& \Psi_0\propto \exp\[-\int_{x_0}^{x}W(y,a_1)dy\]\\
&& \Psi_n^-=\Ad \,(x,a_1)\Psi_{n-1}^-(x,a_2),\ \ n=1,2,3,\dots \er
Therefore, the shape invariance condition actually is an
integrability condition.

\section{the eigenvalues and eigenfunctions of the excited states in first order}
In the following, we rewrite the equations (\ref{3}) in the
Schr\"{o}dinger form. Though the form (\ref{3}) is more familiar
for research, the problem is easier to solve in the original
differential equation than in the equation (\ref{3}). The original
form is obtained from the eq.(\ref{3}) by the transformation \be
x=\cos\t, \ee that is,
\begin{equation}
\left[\frac{1}{\sin \theta}\frac{d}{d\theta}\left(\sin \theta
\frac{d}{d \theta}\right)+ \a \cos ^2 \theta -\frac{m^{2}}{\sin ^2
\theta}\right]\Theta=-E\Theta\label{2}
\end{equation}
the corresponding  boundary conditions become $\T$ is finite at
$\theta=0,\ \pi$,

From eq.(\ref{2}) and by the transformation
\begin{equation}
\Theta =\frac{\Psi}{\sin^{\frac12} \theta}\label{transform}
\end{equation}
we could get the Schr\"{o}dinger form as
\begin{eqnarray}
\frac{d^2\Psi}{d\theta^2}+\left[\frac14+ \a\cos ^2 \theta
 -\frac{m^{2}-\frac14}{\sin ^2
\theta}+A_s\right]\Psi=0\label{main eq}
\end{eqnarray}
and the boundary conditions
\begin{equation}
\Psi|_{\theta=0}=\Psi|_{\theta=\pi}=0.
\end{equation}
The equations (\ref{main eq}) show the potential is
\begin{equation}
V(\theta,\a, m)=-\frac14- \a\cos ^2 \theta
+\frac{m^{2}-\frac14}{\sin ^2 \theta}\label{potential m s b}
\end{equation}
The super-potential $W$ could be determined by
\begin{equation}
W^2-W'=V(\theta,\alpha, m)-E_0=-\frac14+\frac{m^2-\f14}{\sin ^2
\theta}- \alpha\cos ^2 \theta -E_0\label{2potential m=s=0 alpha 0}
\end{equation}
$E_0$ is the the first eigenvalue (the ground state energy) of the
equations (\ref{main eq}). Subtracting $E_0$ to make it possible
to factorize eqns.(\ref{main eq}). Actually, it is difficult to
find the solutions of the equation (\ref{2potential m=s=0 alpha
0}). So the perturbation methods come naturally, which have been
detailed in the reference \cite{tian1} where the perturbation
method in supersymmetry quantum has been used to resolve the
ground state function of the spheroidal equations in approximation
of little parameter $\a$. Here is brief accounting of the results.
One just expands the super-potential $W$, the central concept of
supersymmetry quantum, in the series form of the parameter $\a$.
Because the potential is already in the series of $\a$,  one could
solve the series forms of the super-potential $W$ by solving
 term by term. Also note the fact the ground energy $E_0$ must be
 in the series form of the parameter $\a$ and could be obtained by
 the boundary conditions. So, the super-potential $W$ and $E_0$ could be expanded as series of the
parameter $\a$, that is,
\begin{equation}
W=W_0+\alpha W_1+\alpha ^2 W_2+\alpha ^3 W_3+\ldots .\label{W}
\end{equation}
\be E_0=\sum_{n=0}^{\infty}E_{0n}\alpha^n \ee The perturbation
equation becomes
\begin{equation}
W^2-W'=V(\theta,\alpha,
m)-\sum_{n=0}^{\infty}E_{0n}\alpha^n=-\frac14+\frac{m^2-\f14}{\sin
^2 \theta}- \alpha\cos ^2 \theta
-\sum_{n=0}^{\infty}E_{0n}\alpha^n\label{2potential m=s=0 alpha}
\end{equation}
There are two lower indices in the parameter $E_{0n}$ with the
index $0$ refereing to the ground state and the other index $n$
meaning the nth term in parameter $\a$. The last term
$\sum_{n=0}^{\infty}2E_{on}\alpha^n$ is subtracted from the above
equation in order to make the ground state energy actually zero
for the application of the theory of SUSYQM. Later, one must add
the term to our calculated eigen-energy.

In the reference \cite{tian1}, $W_0,\ W_1,\ W_2$ are
 \br &&W_0=-\(m+\f12\) \cot\theta,\ E_{00}=m(m+1).\label{w-0}
\\&&
W_1=\f{\sin\t\cos\t}{2m+3}\label{good result of w1}\\&&W_2=
\[\f{-\sin\t
\cos\t}{(2m+3)^3(2m+5)}+\f{\sin^3\t \cos\t}{(2m+3)^2(2m+5)}\]
\label{w2 in complex 1}.\er

The ground eigenfunction upon to the second order becomes \br
\Psi_0&=&N\exp\[-\int Wd\theta\]\\&=& N\exp\[-\int W_0d\theta\
-\alpha \int
 W_1d\theta-\a^2\int W_2d\t\]*\exp{O(\a^3)}\\
&=&\left(sin{\theta}\right)^{m+\frac12 } \exp{\[-\frac{\alpha
\sin^2\theta}{4m+ 6}\]}\nonumber\\ &*&  \exp{\[\f{\a^2\sin^2\t
}{2(2m+3)^3(2m+5)}-\f{\a^2\sin^4\t
}{4(2m+3)^2(2m+5)}\]}*\exp{O(\a^3)} .\er

 Because the good result of the quantity $W_1$ in the form of
 the equation (\ref{good result of w1}), the potential upon to
 the first order have the property of the shape-invariance. In
 order to see this property clearly, we rewrite the super-potential $W$ as
 \be W=A_1W_0+\a B_1W_1+O(\a^2)\ee
 then

\br V_-(A_1, B_1, \theta)&=&W^2-W'\nonumber \\
&=&\[(m+\f12)^2A_1^2-(m+\f12)A_1\]\csc^2\theta-\f{B_1\a}{2m+3}\[(2m+1)A_1+2\]
\cos^2\theta\nonumber\\&+& \[\f{B_1\a}{2m+3}-(m+\f12)^2A^2_1\]+O(\alpha^2)\\
V_+(A_1, B_1, \theta)&=&W^2+W'\nonumber\\
&=&\[(m+\f12)^2A_1^2+(m+\f12)A_1\]\csc^2\theta-\f{B_1\a}{2m+3}\[(2m+1)A_1-2\]
\cos^2\theta\nonumber\\&+& \[-\f{B_1\a}{2m+3}-(m+\f12)^2A^2_1\]+O(\alpha^2)\\
&=&V_-(A_2, B_2, \theta)+R(A_1, B_1)
)\nonumber\\
&=&\[(m+\f12)^2A_2^2-(m+\f12)A_2\]\csc^2\theta-\f{B_2\a}{2m+3}\[(2m+1)A_2+2\]
\cos^2\theta\nonumber\\&+& \[\f{B_2\a}{2m+3}-(m+\f12)^2A^2_2\]+O(\alpha^2)\nonumber\\
&&+R(A_1, B_1)+O(\alpha^2) \er where

\be R(A_1,
B_1)=(m+\f12)^2A_2^2-(m+\f12)^2A_1^2-\f{B_2+B_1}{2m+3}\a \ee with
$A_1=B_1=1$. From the above equations, it is easy to get
\br (m+\f12)A_2&=&f_1(A_1)=(m+\f12)A_1+1\\
B_2&=&f_2(A_1,A_2)=\f{(2m+1)A_1-2}{(2m+1)A_1+4}B_1\\
R(A_1, B_1)&=&(m+\f12)A_1+1 -\f{\a}{3}\({B_2+B_1}\)\er Define \br
A_{k+1}&=& f_1(A_{k})=A_k+\f1{m+\f12}\label{a-relation}\\
B_{k+1}&=& f_{2}(A_k,B_k)=\f{(2m+1)A_k-2}{(2m+1)A_k+4}B_k \label{b-relation}\\
R(A_k,B_k,B_{k+1})&=&(2m+1) A_k+1- \f{\a}{3}\({B_{k+1}+B_k}\),\er
then the n+1th energy (or the nth excited energy)upon to the first
order is \br
E^-_n&=&\sum_{k=1}^{n}R(A_k,B_k,B_{k+1})\\ &=& \sum_{k=1}^{n}{\((2m+1)A_k\)-\f{\alpha}{2m+3}\(B_{k+1}+B_k\)}\\
&=& n(2m+n+1)-\f{\a}{2m+3}\[2\(B_1+B_2+\dots
+B_{n}\)-B_1+B_{n+1}\].\er By the relation of
eq.(\ref{b-relation}), we get \br
\(B_{k+1}+B_k\)=\f{2k}{2m+3}\(B_k-B_{k+1}\),\er therefore, \br
\(B_1+B_2+\dots +B_{n}\)=\f14\[(2m+3)B_1-(2n+2m+3)B_{n+1}\].\er We
could get $B_{n+1}$ from the relation (\ref{b-relation}), that is,
\be
B_{n+1}=\f{(2m-1)(2m+1)(2m+3)}{\(2n+2m-1\)\(2n+2m+1\)\(2n+2m+3\)}B_1,\ee
so, \br E_n^-&=&-\f{\a}{2(2m+3)}\[(2m+1)B_1-\(2n+2m+1\)B_{n+1}\]+O(\a^2)\nonumber\\
&=&
n(2m+n+1)\nonumber\\
&-&
\f{\a}{2(2m+3)}\[(2m+1)B_1-\f{(2m-1)(2m+1)(2m+3)}{\(2n+2m-1\)\(2n+2m+3\)}B_1\]+O(\a^2).\er
As stated in the reference \cite{tian1}, before, the last term in
eq.(\ref{2potential m=s=0 alpha}) must be added to $E_n^-$ for
actual calculation, that is, \br &&E_n^-+E_{00}+E_{01}\a
=n(2m+n+1)+m(m+1)-\f{\a}{2m+3}-\nonumber\\
&-&\f{\a}{2(2m+3)}\[(2m+1)B_1-\f{(2m-1)(2m+1)(2m+3)}{\(2n+2m-1\)\(2n+2m+3\)}B_1\]
+O(\a^2)\\
&=&
n(2m+n+1)+m(m+1)-\f{\a}2\[1-\f{(2m-1)(2m+1)}{\(2n+2m-1\)\(2n+2m+3\)}\].\er

Notice that our formula is different  from that of the reference
\cite{flammer}, the relation is that our $m+n$ is equivalent to
$l$ in the reference \cite{flammer}, that is, $n=l-m$  and an
overall negative sign between them. Then we can compare the
results, which is same: \br
E_n^-+E_{00}+E_{01}\a+0(\a)&=&(l-m)(l+m+1)+m(m+1)-\f{\a}2\[1-\f{(2m-1)(2m+1)}{\(2l-1\)\(2l+3\)}\]\nonumber
\\&=& l(l+1)-\f{\a}2\[1-\f{(2m-1)(2m+1)}{\(2l-1\)\(2l+3\)}\].\er

The nth excited state upon to the first order is obtained by \br
&& \psi_n^-(\theta,A_1,B_1)= \Ad \psi_{n-1}^-(\theta,A_2,B_2)\\&
\propto & \Ad (\theta, A_1,B_1)\Ad (\theta, A_2,B_2)\dots \Ad
(\theta, A_n,B_n)\psi_{0}^-(\theta,A_{n+1},B_{n+1}),\er with \be
\psi_{0}^-(\theta,A_{n+1},B_{n+1})=\left(sin{\theta}\right)^{(m+\frac12)A_{n+1}}
\exp\[\left(-\frac{\alpha B_{n+1}
\sin^2\theta}{4m+6}\right)\]*\exp{O(\a^2)},\ee and \be \Ad
(\theta, A_n,B_n)=-\f d{d\theta}-(m+\f12)A_n\cot\theta
-\f{\a}{4m+6}B_n\sin2\theta \ee where $A_n,\ A_{n+1}$ satisfy the
relation of (\ref{a-relation}),(\ref{b-relation}).

\section{conclusion and discussion}
In the paper, the first order term of the super-potential obtained
has the good property of shape-invariance by the formula of
(\ref{good result of w1}), this in turn makes it easy to get the
pleasant results of the excited energy and the corresponding
excited state functions from the counterparts of the ground state.
The results are much more like case of the associated-Lengdre
functions. They are new and interesting. The method can also be
used to solve higher order terms of the super-potential and the
related super-invariance problem of the spin-weighted spheroidal
functions. The methods apply  to the spin-weighted spheroidal
functions for the case $s\ne 0$. These cases have been studied too
and will be reported.
\section*{Acknowledgements}
We greatly appreciate the hospitality of the Gravitation Theory
group and the MCFP in UMD . This work was supported in part by the
National Science Foundation of China  under grants No.10875018,
No.10773002.

\end{document}